\documentclass[10pt]{bmc_article}

\usepackage{cite} 
\usepackage{url}  
\usepackage{ifthen}  
\usepackage{multicol}   
\usepackage[utf8]{inputenc} 
\usepackage{amssymb}
\usepackage{textcomp}
\usepackage{tipa}
\usepackage{wasysym}
\usepackage{graphicx}
\newboolean{publ}

\newenvironment{bmcformat}{\begin{raggedright}\baselineskip20pt\sloppy\setboolean{publ}{false}}{\end{raggedright}\baselineskip20pt\sloppy}

\begin{document}
\begin{bmcformat}


\title{Ge quantum dot arrays grown by ultrahigh vacuum molecular beam epitaxy 
on the Si(001) surface: nucleation, morphology and CMOS compatibility
}
 

\author{Vladimir A Yuryev\correspondingauthor${~}$%
       \email{V A Yuryev
- vyuryev@kapella.gpi.ru}%
and
         Larisa V Arapkina\correspondingauthor%
         \email{L V Arapkina - arapkina@kapella.gpi.ru}%
      }


\address{%
A. M. Prokhorov General Physics Institute of RAS, 38 Vavilov Street, Moscow 119991, Russia
}%

\maketitle


\begin{abstract}
      


Issues of morphology, nucleation and growth of Ge cluster arrays deposited by ultrahigh vacuum molecular beam epitaxy  on the Si(001) surface are considered. 
Difference in nucleation of quantum dots   during Ge deposition at  low ($\apprle$\,600{\textcelsius}) and high ($\apprge$\,600{\textcelsius}) temperatures is studied by high resolution scanning tunneling microscopy. 
 The atomic models of growth of  both species of Ge huts---pyramids and wedges---are proposed. The growth cycle of Ge QD arrays at low temperatures is explored.
 A problem of lowering of the array formation temperature is discussed with the focus on CMOS compatibility of the entire process; a special attention is paid upon approaches to reduction of treatment temperature during the Si(001) surface pre-growth cleaning, which is at once a key and the highest-temperature phase of the Ge/Si(001) quantum dot dense array formation process. The temperature of the Si clean surface preparation, the final high-temperature step of which is, as a rule, carried out directly in the MBE chamber just before the structure deposition, determines the compatibility of formation process of Ge-QD-array based devices with the CMOS manufacturing cycle. Silicon surface hydrogenation at the final stage of its wet chemical etching during the preliminary cleaning is proposed as a possible way of efficient reduction of the Si wafer pre-growth annealing temperature.

\end{abstract}

\ifthenelse{\boolean{publ}}{\begin{multicols}{2}}{}


\section*{Introduction: Background and problem statement\pb}

Heteroepitaxial structures Ge/Si and SiGe/Si are among the most promising materials of modern nanoelectronics and nanophotonics \cite{Wang-Cha,Wang-properties,QDFET,WG-near_IR,Photonic_crystal,QDIP-MIS,QDIP-Wang2,QDIP-Wang1,QDIP-Wang,Dvur_book_total,Dvur-IR-20mcm,Pchel_ordered_dots}.
Lately industry has developed numerous radiofrequency  devices on the basis of SiGe/Si structures with bands wider than 100\,GHz, which already compete with GaAs-based components. Except that, SiGe-based technology has allowed one to approach to development of the most important elements of single-crystalline integrated microphotonics---laser diodes and detectors for  fiberoptic communications; SiGe-based waveguides are already available. So, the forecasted forthcoming breakthrough would enable the solution of two important problems---(i) development of monolithic VLSI circuits for fiberoptic telecommunications and (ii) replacement of electronic data buses by optical ones. Additionally, encouraging  results have been recently obtained in development of emitting THz and mid-infrared devices based on SiGe/Si-heterostructures. And finally, application of Ge/Si and SiGe/Si hetersructures might enable a breaking progress in IR imaging technology opening a way to creation of multispectral photodetector arrays integrated with readout circuitry on a single-crystalline chip. 
\pb

Dense arrays of Ge quantum dots (QD) are of  importance to all practically significant applications in optoelectronics and microelectronics. QD array is usually referred to as dense array if interactions among adjacent clusters play an important role \cite{Yakimov}, i.e., a tunnel coupling between Ge clusters arise \cite{Ge-Si-QD_Coulomb-correlations};  such arrays should be considered as a whole in terms of behaviour of current carrier and transport properties \cite{Dvur_book_total}. Ge/Si hetrostructure can include both isolated QD arrays, i.e., arrays which do not mutually interact and separated by a thick enough ($>$\,30\,nm) Si layer, and superlattices of QD arrays, in which arrays are separated by thin Si layers and which, like SiGe/Si superlattices with quantum wells, represent  a single coupled system. In addition, it is known that, if distances between layers of Ge clusters  are  as small as unities of nanometers in the direction of the structure growth, ordering of  Ge clusters is observed in this direction \cite{QD_superlattices}; they form chains which can be composed by dozens of clusters if a number of layers is large \cite{Ge-Si-QD_molecules}. Like atoms, Ge clusters form a sort of molecules in which electron density redistributes among clusters depending on distances between them (thicknesses of Si buffer layers), as if changing a type of chemical bond from covalent to ionic. This phenomenon opens a wide perspective   to designing heterostructures with various optical and electrical properties. 
\pb

Recently, an interest of researchers has been attracter by heterostructures with ordered arrays of quantum dots, in particular, by ordered arrays of Ge nanoclusters in the Si matrix \cite{Ge_QD_crystal}. An idea of formation of a QD array, which would combine advantages  of a single QD with benefits of a dense array, seems to be the most promising.$^{\rm a}$ Controllable ordering of clusters  in all three directions would enable creation of a volumetric crystal in which QDs play a role of atoms. In such artificial crystal, unique opportunities appear to designing wave functions of carriers by filling corresponding quantum states of QDs by electrons or holes (like it is the case for $s$, $p$ and $d$ atomic configurations). As opposed to impurity states in semiconductors, dense 3D array of QDs would be an ensemble of multicharged centres in which an essential role would be played by the Coulomb potential. A concept of the QD crystal, which is considered as a 3D lattice of artificial atoms, implies a new material with spatial ordering on the scale comparable with the de~Broglie wave-length for electrons.  Non-locality of the quantum-mechanical bonding together with the Coulomb interaction of carriers localized in close QDs may result in new optical and electronic properties arising from the collective nature of electronic states.   In contrast to stochastically located QDs, in this case these properties would not be averaged over components of a crystal. Main properties of such ensemble would reproduce peculiarities inherent to ordinary solids, such as appearance of two-dimentional or three-dimentional minibands, separated by minigaps, in lieu of localized quantum states intrinsic to separate QDs. It is necessary for QD crystal that QDs would be ordered to precise periodicity, the sizes of QDs would equal, and distances  between QDs would be small enough for wave functions to overlap.
\pb

Such QD crystals would be very prospective for application in nanoelectronics, spintronics and, likely, in quantum computing, as well as in devices of silicon optoelectronics such as highly efficient sources and detectors of infrared and terahertz emission enabling integration  to silicon VLSI circuitry.
\pb

Main restriction for use of such Ge/Si heterostructures with dence arrays of self arranged Ge QDs  are associated with the spread sizes of Ge clusters and their tend to disordering on the surface. Both these factors cause tailing of a discrete spectrum. 
Additional difficulty is the necessity for all technological steps to be embedded into VLSI manufacturing process or, in other words, 
meet requirements of CMOS compatibility.
\pb

To be able to solve the above ambitious task, a deep knowledge of physical processes on silicon surface during its preparation and in germanium and silicon films during the heterostructure formation is strongly required. This article represents some results of our resent investigations in this direction.

\section*{Methods\pb}
  \subsection*{Equipment\pb}    

The experiments were carried out using an integrated ultrahigh vacuum
instrument \cite{classification,stm-rheed-EMRS,CMOS-compatible-EMRS} built on the basis of the Riber~SSC\,2
surface science center with the  EVA\,32 molecular-beam epitaxy (MBE) chamber  equipped with the Staib Instruments RH20 diffractometer of
reflected high-energy electrons (RHEED) and connected through a transfer line to the
STM GPI-300 ultrahigh vacuum scanning tunnelling microscope
\cite{gpi300,STM_GPI-Proc,STM_calibration}.$^{\rm b}$ A preliminary
annealing and outgassing chamber is also available in the instrument.
\pb

The pressure of about $5\times 10^{-9}$\,Torr was kept in the preliminary
annealing chamber. 
The MBE chamber was evacuated down to about $10^{-11}$\,Torr before processes; the pressure grew to nearly $2\times 10^{-9}$\,Torr at most during the sample surface deoxidization process and $10^{-9}$\,Torr during Ge or Si deposition.   
The residual gas pressure did not exceed $10^{-10}$\,Torr in the STM chamber. 
\pb

The instrument enables the STM study of samples at any stage of Si surface preparation and MBE growth. The samples can be consecutively moved into the STM chamber
for the analysis and back into the MBE vessel for further treatments or Ge, Si or SiGe deposition
as many times as required never leaving the UHV ambient and preserving the required cleanness for MBE growth and STM investigations with atomic resolution. RHEED
experiments can be carried out {\it in situ}, i.e., directly in the
MBE chamber during a process \cite{stm-rheed-EMRS}. 
\pb

Sources with the electron beam evaporation were used for Ge or Si deposition.
 The deposition rate and  coverage were measured using
the Inficon Leybold-Heraeus  XTC\,751-001-G1 film thickness monitor equipped with the graduated
in-advance quartz sensors installed in the MBE chamber. 
Tantalum radiators were used  for sample heating from the rear side in both
preliminary annealing and MBE chambers. The temperature was
monitored with chromel--alumel and tungsten--rhenium thermocouples of the heaters
in the preliminary annealing and MBE chambers, respectively. The
thermocouples were mounted in vacuum near the rear side of the
samples and {\it in situ} graduated beforehand against the
IMPAC~IS\,12-Si pyrometer which measured the sample temperature  through chamber windows. The temperature distribution uniformity over a surface was also investigated in advance; the deviations from mean values were found to be within {$\pm$3{\textcelsius}} for the  half-radius areas around the centers of  $2^{\prime\prime}$ wafers over the whole temperature interval applied in this study.
\pb

The  composition of residual atmosphere in the MBE
camber was monitored using the SRS~RGA-200 residual gas analyzer
before and during the process.
\pb

The STM tip was {\it ex situ} made of the tungsten
wire and cleaned by ion bombardment \cite{W-tip} in a special UHV
chamber connected to the STM one. 
\pb

In this work, the images were obtained in the
constant tunnelling current ($I_{\rm t}$) mode at the room
temperature. The STM tip was zero-biased while the sample was
positively or negatively biased ($U_{\rm s}$) when scanned in empty-
or filled-states imaging mode.
\pb

Original firmware \cite{gpi300,STM_GPI-Proc,STM_calibration} was
used for data acquisition; the STM images were processed afterward
using the WSxM software \cite{WSxM}.

\subsection*{Sample preparation procedures\pb}  

\subsubsection*{Preparation of samples with deposited Ge layers}

Initial samples for STM were  $8\times 8$~mm$^{2}$ squares cut from the
specially treated commercial boron-doped    Czochralski-grown (CZ) Si$(100)$ wafers
($p$-type,  $\rho\,= 12~{\Omega}\,$cm).  After washing and chemical treatment
following the standard procedure described elsewhere
\cite{etching} (which included washing in
ethanol, etching in the mixture of HNO$_3$ and HF and rinsing in the
deionized water \cite{cleaning_handbook}), the silicon substrates were mounted on the
molybdenum STM holders and inflexibly clamped with the tantalum
fasteners. The STM holders were placed in the holders for MBE made
of molybdenum with tantalum inserts. 
Then, the substrates were
loaded into the airlock and transferred into the preliminary
annealing chamber where they were outgassed at the temperature of
around {565{\textcelsius}} for more than 6\,h. After that, the substrates were
moved for final treatment and Ge deposition into the MBE chamber 
where they were subjected to 
two-stages annealing during heating with stoppages at {600{\textcelsius}}
for 5\,min and at {800{\textcelsius}} for 3\,min \cite{classification}. The
final annealing at the temperature greater than {900{\textcelsius}} was
carried out for nearly 2.5\,min with the maximum temperature of
about {925{\textcelsius}} (1.5\,min). Then, the temperature was rapidly
lowered to about {750{\textcelsius}}. The rate of the further cooling was
around 0.4{\textcelsius}/s that corresponded to the `quenching' mode
applied in \cite{stm-rheed-EMRS}. The
surfaces of the silicon substrates were completely purified of the
oxide film as a result of this treatment \cite{phase_transition,stm-rheed-EMRS,our_Si(001)_en}.
\pb
 
Ge was deposited directly on the deoxidized Si(001) surface. The
deposition rate was varied from about $0.1$ to $0.15$\,\r{A}/s; the effective Ge film
thickness $(h_{\rm Ge})$ was varied from 3 to 18\,\r{A} for different
samples.  The substrate
temperature during Ge deposition $(T_{\rm gr})$ was {360{\textcelsius}} or {530{\textcelsius}} for the low-temperature mode   and 600 or {650{\textcelsius}} for the  high-temperature mode.
The rate of the sample cooling down to the room temperature was approximately
0.4{\textcelsius}/s after the deposition.
\pb

After  cooling, the prepared samples with Ge layers were moved for
analysis into the STM chamber.  
\pb

\subsubsection*{Preparation of samples for Si(001) surface analysis}

Wafers for Si(001) surface analysis by STM and RHEED were the same as for Ge MBE. Initially, the specimens were chemically etched in the RCA etchant \cite{RCA}  and processed to form a surface  terminated by hydrogen atoms (Si:H). The hydrogenated Si:H samples were prepared by etching in solutions containing HF at the final stage of the RCA process \cite{Nanophysics-2011_H}. We used the following solutions with  pH\,=\,2, 4 or 7: a dilute HF solution (5\% or 0.5\%), buffered NH$_4$F\,+\,HF or NH$_4$F solutions. 
After that, the Si:H samples were pre-treated for 2 hours at the temperature of {$\sim$300{\textcelsius}} and the pressure of less than $5\times 10^{-11}$~Torr in the MBE chamber.
The second phase of the thermal treatment was conducted at the temperatures of 800, 650, 610, 570, 550, 530 or 470{{\textcelsius}}. Duration of this phase was chosen to a form of the RHEED pattern.
The samples were quenched after heat treatments   at the rate of $\sim$0.4 {\textcelsius}/s\cite{stm-rheed-EMRS}.

\section*{Results and Discussion\pb}

\subsection*{Nucleation\pb}

\subsubsection*{Hut nucleation at low and high temperatures}

{\it Nucleation of Ge clusters at low temperatures} has been an issue of numerous experimental and theoretical investigations for a number of years (see a brief review section in article \cite{classification}). 
Recently, we have described two characteristic formations composed by epitaxially oriented Ge dimer pairs and  chains of four dimers on the wetting layer (WL) patches which were interpreted by us as two types of hut nuclei: an individual type for each species of huts---pyramids or wedges (Figure~\ref{fig:nuclei}a-c) \cite{Hut_nucleation,CMOS-compatible-EMRS,classification}. 
These nuclei are always observed to arise on sufficiently large WL patches: there must be enough room for a nucleus on a single patch; a nucleus cannot be housed on more than one patch \cite{Hut_nucleation}.
\pb 

Both types of the hut nuclei appeared to arise at the same WL $M\times N$ patch thickness \cite{CMOS-compatible-EMRS,Array_Initial_Phase}, hence, at the same WL stress to relieve it. Therefore, they appear at the same strain energy (and with equal likelihoods, see Refs.~\cite{classification,Hut_nucleation}). 
This means that they are degenerate by the formation energy: 
if they had different formation energies they would appear at different WL thicknesses; the first of the types of huts, which nucleates on the surface, releases the stress;  the second one never appears therefore. Hence, they can occur only simultaneously and their formation energies can  be only equal.  
\pb

Presently we have no satisfactory explanation of this phenomenon and can only propose a very preliminary interpretation of the observed simultaneous appearance of the two kinds of nuclei on WL. The explanation is based on  modeling of Ge cluster formation energy performed in Ref.~\cite{Domes_first}. Brehm {et al.} \cite{Domes_first} have explored Ge island nucleation during MBE  at much higher temperatures than those applied in this work, therefore  theoretical results of Ref.~\cite{Domes_first} describe the experimental data obtained for the  case of the high-temperature growth mode, which differs considerably from the low-temperature one \cite{classification}. However, the modeling could also apply  for the low-temperature growth. According to Ref.~\cite{Domes_first}, flat Ge islands---in our case, nuclei and small huts---likely occur on WL because of an energy benefit which arises in exposing the compressed \{105\} facets, rather than in relaxing the volumetric elastic energy, as it takes place in the usual Stranski-Krastanov mechanism. At low temperatures, this effect may stabilize clusters, however preventing their further ripening (this agrees with our observations presented recently in Ref.~\cite{classification}). If this is the case, the actual volumetric and structural form of clusters likely do not impact very much in their formation energy.$^{\rm c}$ 
\pb

As distinct from the low-temperature mode, {\it Ge cluster  nucleation  at high temperatures} may go on in two ways. The first way is similar to the process of hut nucleation at low temperatures. Pyramids were observed to nucleate in such a way. Figure~\ref{fig:nuclei}d,e illustrates this statement: the pyramid nuclei, absolutely the same as those observed in the samples grown at low temperature, are seen on the WL patches in the images of the samples obtained at $T_{\rm gr} =$ 650{{\textcelsius}}. Their density was small, and they were mainly situated in the vicinity to large mature pyramids, which arise at early stages of Ge deposition and have much greater sizes than huts formed at low temperatures at the same values of $h_{\rm Ge}$ \cite{Nanophysics-2011_Ge}. The WL surface mainly consisted of  monoatomic steps and narrow terraces in these ares (Figure~\ref{fig:nuclei}d).
\pb

The second way, somewhat resembling the process described by Goldfarb {et al.} \cite{Nucleation} for the case of the gas-source MBE (and thick hydrogenated WL),  is illustrated by Figure~\ref{fig:heap}. At small values of $h_{\rm Ge}$, regions containing excess of Ge atoms were observed on the surface. Usually, they  were not resolved as structured formations and resembled shapeless heaps of Ge (Figure~\ref{fig:heap}a).  Pits usually accompanied them. Heap density was about $10^9$\,cm$^{-2}$. Some of heaps had started to form the $\{$105$\}$ facets during Ge deposition (Figure~\ref{fig:heap}b). 
\pb

Stoppage of Ge deposition and subsequent annealing at $T_{\rm gr}$ resulted in formation of volumetric structures partially or even completely faceted by $\{$105$\}$ planes,    transforming `heaps' to some similarities of huts (Figure~\ref{fig:heap}c,d,e). 
\pb

We have never  observed such process at low temperatures of growth and suppose it to be inherent only to the high-temperature array formation mode.

\subsubsection*{Array nucleation and growth outset}

Since the pioneering work by Mo et al. \cite{Mo}, it has been known that deposition of Ge on Si(001) beyond 3\,ML (1\,ML\,$\approx$\,1.4\,\AA) leads to formation of huts \cite{Mo,LeGoues_Copel_Tromp,Chem_Rev} on WL with high number density ($\apprge 10^{10}$\,cm$^{-2}$, Refs.~\cite{Jesson_Kastner_Voigt,classification,CMOS-compatible-EMRS}). Some later the value of Ge coverage, at which 3D clusters emerged, was confirmed by Iwawaki et al.\cite{Iwawaki_initial} who, in the course of a comprehensive STM study of the low-temperature epitaxial growth of Ge on Si(001) \cite{Iwawaki_initial,Iwawaki_SSL,Iwawaki_dimers,Nishikawa_105},  directly observed appearance of minute (a few ML height) 3D Ge islands at 300{{\textcelsius}} on  $(M\times N)$-patched  WL; deposition of 4\,ML of Ge resulted in formation of a dense array of small huts. 
Various values of Ge coverage, at which the transition from 2D to 3D growth  occurs, are presented in the literature. For example, an abrupt increase in hut density at the coverage of 3.16\,ML was detected  for Ge deposition at {300{\textcelsius}} and 0.06\,ML/min \cite{Jesson_Kastner_Voigt}. A detailed phase diagram of the Ge film on Si(001) derived from experiments carried out by recording RHEED gave the coverages corresponding to the ``2D-to-hut'' transition from $\sim$2.5 to $\sim$3\,ML for the growth temperature interval from 300 to 400{{\textcelsius}} (and different values for different temperatures) \cite{Nikiforov_Gi-Si_RHEED}. Photoluminescence study of Ge huts deposited at the temperature of 360{{\textcelsius}} showed that evolution from ``quantum-well-like'' (attributed to WL) to ``quantum-dot-like'' (attributed to Ge huts) emission occurred at a coverage of $\sim$4.7\,ML in  PL spectra obtained at 8\,K \cite{PL_ultrasmall_Ge}. Hut formation studied by high resolution low-energy electron diffraction and surface-stress-induced optical deflection  evidenced that at deposition temperature of 500{{\textcelsius}} hut formation suddenly set in at a coverage of 3.5\,ML \cite{LEED_Ge-nucleation}. And finally, for theoretical studies the WL thickness and consequently the hut formation coverage is usually assumed  to equal 3\,ML \cite{hut_stability}. As it is seen from the above examples,  there is no unambiguous information presently about the coverage at which huts arise or, more accurately, about the thickness of the WL $M\times N$  patch on which a cluster nucleate during Ge deposition. STM studies show the WL thickness to equal 3\,ML only on the average: $M\times N$ patches have slightly different thicknesses (\textpm\,1\,ML) around this value \cite{Iwawaki_initial,classification,CMOS-compatible-EMRS,Hut_nucleation,atomic_structure}.
In this section, we determine by means of high resolution STM an accurate value of the $M\times N$ patch height at which hut nucleate at 360{{\textcelsius}}.
\pb

Figure~\ref{fig:4,4A_5,1A}a demonstrates a typical STM micrograph of the  ($M\times N$)-patched WL ($h_{\rm Ge}=4.4$~\r{A}, $\sim$3.1\,ML). This image does not demonstrate any feature which might be recognized as a hut nucleus (Figure~\ref{fig:nuclei}a-c)  \cite{Hut_nucleation}. Such features first arise at the  coverages of $\sim$5\,\r{A}: they are clearly seen in Figures~\ref{fig:4,4A_5,1A}b-d, which demonstrate a moment when the array have just nucleated ($h_{\rm Ge}=5.1$\,\r{A}, $\sim$3.6\,ML). However, we succeeded to find minute pyramid and wedge at this $h_{\rm Ge}$ (Figure~\ref{fig:4,4A_5,1A}d)---both as small as 2\,ML over the patch surface (we measure cluster heighs from  patch tops)---which indicate that hut nucleation had started a little earlier. 
\pb

It can be concluded from these observations that hut arrays nucleate at a coverage of $\sim$5.1\,\r{A} ($\sim$3.6\,ML) when approximately a half of patches are as thick as 4\,ML. We can suppose then that huts nucleate on those patches whose thickness have reached (or even have exceeded) 4\,ML. 
\pb

\subsection*{Morphology\pb}

\subsubsection*{Wetting layer reconstruction}

Evolution of WL patches during MBE is illustrated by Fig.~\ref{fig:WL_patches}. In full agreement with the data of Ref.\,\cite{Iwawaki_initial}, both $c(4\times 2)$ and $p(2\times 2)$ reconstructions are observed on tops of the $M\times N$ patches in all images except for the image given in Figure~\ref{fig:WL_patches}a ($h_{\rm Ge}=4.4$~\r{A}) in which only the $c(4\times 2)$ structure is recognized. A magnified image of the $p(2\times 2)$ structure illustrating its characteristic zig-zagged shape and resolving separate upper atoms of buckled Ge dimers is given in Figure~\ref{fig:WL_patches}d.
\pb

Formation of a hut nucleus on a patch reconstructs its surface; a new formation changes the structure of the topmost layer  to that specific for a particular type of nuclei, in the present case, to the structure of the pyramidal hut nucleus (Figure~\ref{fig:WL_patches}e). However the residual $c(4\times 2)$ structure  still remains on the lower terrace of the patch. At the same time, the $p(2\times 2)$ structure stays on the top of the adjacent patch \cite{Array_Initial_Phase}. 
\pb

We can conclude now that $c(4\times 2)$ and  $p(2\times 2)$ surface structures occurring on the $M\times N$ patches should be energetically degenerate. In addition, the above observation rises a question if there is some connection between the form of a patch top reconstruction and a species of hut which could nucleate on it or, in other words, whether the patch top reconstruction controls hut nucleation and determines its species.
\pb

\subsubsection*{Growth and structure of pyramids and wedges}

Let us consider possible scenarios of hut growth after nucleation on WL. Earlier \cite{classification,atomic_structure,CMOS-compatible-EMRS}, we have already proposed structural models of both species of huts and briefly discussed processes and atomic models of their formation giving a few drawings with identical apexes as examples and allowing the readers to construct the missing structural schemes. However, crystallography  allows one to arrive to two different solutions for wedge-like huts, and  additional empirical knowledge and STM data is required to discriminate  between them. Both solutions are given in Figure~\ref{fig:growth}. The first scenario of growth  assumes {\it uniform addition of Ge atoms} to all four facets of huts (follow a series number I in Figure~\ref{fig:growth}). In this case wedge-like huts have different ridge structures  (the ridge width and location of atoms on it) depending on cluster height. The initial ridge structure, which form on top of 2-ML wedge reconstructing the nucleus \cite{Hut_nucleation,CMOS-compatible-EMRS},  should then occur on the ridge every 5\,ML over the nucleus, i.e., only the wedges  of 2, 6, 11, etc. ML height over WL can have the same ridge structure. This contradicts our observations according to which the structure of hut apexes always remains and depends on only the hut species. Therefore, we are made to come to a different solution in which {\it Ge atoms are added to  facets non-uniformly} (see series II for wedges and pyramids in Figure~\ref{fig:growth}). In this scenario, the structures of both apexes of huts are independent of cluster heights, that agrees with experimental observations. 
\pb

\subsubsection*{Complete cycle of Ge QD array growth at low temperatures}

The STM images of the surfaces of the germanium layers grown at $T_{\rm gr} =360${{\textcelsius}} with various $h_{\rm Ge}$ values are shown in Figure~\ref{fig:array_life}  where the evolution of the Ge layer on the Si(001) surface in the process of low-temperature MBE is seen. Hut clusters on the Ge surface have not nucleated at $h_{\rm Ge} = 4.4$\,\r{A}, and the STM image in Figure~\ref{fig:array_life}a exhibits only the well-known structure of the wetting layer with the $c(4 \times 2)$ or $p(2 \times 2)$ reconstruction inside $M \times N$ blocks \cite{CMOS-compatible-EMRS,Hut_nucleation}. 
The array nucleates at $h_{\rm Ge} \sim 5$\,\r{A} (Figure~\ref{fig:4,4A_5,1A}) but 3D huts mainly form at higher coverages \cite{CMOS-compatible-EMRS,Array_Initial_Phase}. 
Hut arrays initially evolve with increasing  $h_{\rm Ge}$
by  concurrent growth of available clusters and nucleation of new ones resulting in progressive rise of  hut number  density.
Huts are clearly seen in Figure~\ref{fig:array_life}b for $h_{\rm Ge} = 6$\,\r{A}; their density and sizes increase; the number density of huts reaches maximum at $h_{\rm Ge} = 8$\,\r{A} (Figure~\ref{fig:array_life}c); clusters with various sizes---completely formed clusters, recently nucleated small clusters, and nuclei with a height of 1\,ML over the Ge WL---are simultaneously present on the surface \cite{CMOS-compatible-EMRS,Hut_nucleation,Array_Initial_Phase}. This array is very inhomogeneous both in the sizes of the clusters and in composition; it includes regular pyramidal and elongated wedge-shaped clusters, but wedge-shaped clusters with a large spread in the lengths dominate \cite{classification}. The array is most homogeneous at $h_{\rm Ge} = 10$\,\r{A}  (Figure~\ref{fig:array_life}d) \cite{our_Raman_en}, clusters cover almost the entire surface of the wetting layer, the fraction of small clusters decreases noticeably, and large clusters begin to coalesce. At $h_{\rm Ge} = 14$\,\r{A}, most clusters coalesce near their bases (Figure~\ref{fig:array_life}e,f), and the free wetting layer almost disappears from the field of view of STM, but the array consists of individual clusters. At $h_{\rm Ge} = 15$\,\r{A}, the coalescence of clusters continues and a transition to the growth of a two-dimensional film of nanocrystalline germanium begins (Figure~\ref{fig:array_life}g). Nevertheless,
the hut nucleation  continues on small lawns of WL rarely preserved, surrounded by large huts, even at
as high coverages as 15\,\r{A}, when virtually total coalescence of
the mature huts have already happened \cite{CMOS-compatible-EMRS}.
Finally, at $h_{\rm Ge} = 18$\,\r{A}, it is seen that the array of Ge clusters disappears and although the roughness of the surface is still pronounced, the Ge layer grows as a continuous nanocrystalline film (Figure~\ref{fig:array_life}h). A chaotic conglomeration of faceted hillocks and pits composes the film; steep facets appear around the pits (Figure~\ref{fig:array_life}i). However,  Ge WL $(M\times N)$-patched structure is clearly resolved on the bottom of pits and WL lawns (Figure~\ref{fig:array_life}j,k). WL appears to be a very stable formation. 
\pb

The density of the wedge-shaped clusters increases when $h_{\rm Ge}$ increases up to 8\,{\AA} and, then, decreases slowly, whereas the density of the pyramid-shaped clusters decreases exponentially in the process of growth of the array \cite{classification,CMOS-compatible-EMRS}. The total density of the clusters is about $3.5 \times 10^{11}$, $5.8 \times 10^{11}$, $5.1 \times 10^{11}$, and $2.3 \times 10^{11}$~cm$^{-2}$ at $h_{\rm Ge} =$ 6, 8, 10, and 14\,\r{A}, respectively. 
From the capacitance-voltage characteristics of the samples with Ge/Si(001) heterostructures, the surface densities of holes in them were earlier estimated as $3.4  \times 10^{11}$, $7.0  \times 10^{11}$, and $1.7 \times 10^{11}$~cm$^{-2}$ for  $h_{\rm Ge} =$ 6, 10, and 14\,\r{A}, respectively. These values almost coincide with the densities of the Ge clusters in arrays \cite{C-V_ICDS-25,our_THz_en}. Notice also, that very high terahertz conductivity was observed  by Zhukova et al. \cite{our_THz_en} in the samples with $h_{\rm Ge} =$ 8, 9, 10, and 14\,\r{A}   which drastically decreased for  $h_{\rm Ge} =$ 18\,\r{A} and was not detected at all at 6\,\r{A} and lower values of $h_{\rm Ge}$.
\pb

\subsection*{CMOS compatibility\pb}


CMOS compatibility of technological processes based on Ge/Si heteroepitaxy imposes a hard constraint on conditions of all the phases of the heterostructure formation including  Si wafer thermal cleaning and surface preparation to epitaxial growth. Formation of a device structure with QD arrays as a rule must be one of the latest operations of the whole device production cycle because otherwise the  QD arrays would be destroyed by
further high-temperature annealings. High-temperature processes during Ge/Si heterostructure formation on the
late phase of  chip production would, in turn, certainly wreck a
circuit  already formed on the crystal. Therefore, lowering of the
array formation temperature down to the values of $\apprle 450${{\textcelsius}}, as well as decreasing of the wafer annealing temperatures
and times during the clean Si(001) surface preparation, is strongly required
\cite{CMOS-compatible-EMRS}. We refer to the Ge QD arrays and heterostructures based on them which satisfy this requirement as CMOS-compatible.
\pb

\subsubsection*{Si(001) hydrogenation  as a promising way of reduction of the surface cleaning temperature}


Development of a procedure of clean Si(001) surface preparation  at lowered temperatures and/or by short thermal treatments is  a keystone of creation of a
CMOS compatible process  of nanoelectronic VLSI fabrication \cite{classification,Hut_nucleation}. One of the ways of solving this problem is  surface hydrogenation  during wet chemical etching with subsequent  hydrogen desorption from the surface in UHV ambient \cite{cleaning_handbook,thermal_desorption}. In this connection, an issue of surface structure after these treatments becomes a task of primary importance taking into account a possible effect of  Si surface atomic-scale roughness on  formation of nanostructured elements (e.\,g., self-assembled Ge quantum dot nucleation on wetting layer in Ge/Si(001) heterostructures \cite{Hut_nucleation,CMOS-compatible-EMRS,Ion_irradiation,Smagina}).
In this section, we present data of our recent investigations conducted by means of STM and RHEED on preparation of  clean Si(001) surfaces  by  hydrogenation and thermal desorption of hydrogen in an UHV MBE chamber after wet chemical etching by the RCA process \cite{Nanophysics-2011_H}. 
\pb

It is known \cite{H-desorption_Sakamoto,thermal_desorption} that the temperature of surface cleaning depends on composition of etchants used for hydrogenation.
 Solutions based on HF, with pH varied from 2 to 7, are typically used  for surface hydrogenation. A number of silicon hydrides form on the Si(001) surface by the reaction of hydrogen with Si, and the most typical ones are monohydride and dihydride (Figure~\ref{fig:SiH_schematic}). A fraction of dihydride on the surface grows with the 
increase of pH; monohydride desorbs from the surface at higher temperature \cite{thermal_desorption}.
\pb

The main results of our studies are as follows.
\pb

Explorations of {\it hydrogenated surfaces} \cite{Nanophysics-2011_H} have evidenced that regardless of the type of solution used for surface hydrogenation, RHEED patterns correspond with unreconstructed $1\times 1$ surface (Figure~\ref{fig:SiH}).  Broad streaks  with pronounced 3D-related structure form the RHEED patterns for the samples etched in HF solutions (Figure~\ref{fig:SiH}a,b); high intensity of the Kikuchi lines indicates that the surface is  highly smooth and ordered on macroscopic scale. Visible local enhancement of signal of the RHEED patterns takes place  owing to overlapping of Kikuchi lines. 
The shapes of the streaks corresponding to the surface well developed  on the monoatomic scale (3D spots) are detected in the patterns of the samples treated in NH$_4$F  solutions (Figure~\ref{fig:SiH}d,e). According to STM data, the surface is more rough in the case of etching in the  NH$_4$F  solutions than in the case of hydrogenation in solutions based on HF. 
\pb

STM measurements have shown that clean Si(001) surfaces  may be obtained as a result of {\it hydrogen thermal desorption} in the interval from 470 to 650{{\textcelsius}} \cite{Nanophysics-2011_H}, but their roughness depends on chemical treatment applied for hydrogenation and temperature of subsequent annealing (Figures~\ref{fig:HF} and \ref{fig:SiH4F}). If dilute HF solutions and annealings of moderate duration at temperatures higher than 600{{\textcelsius}}   are applied, smooth surfaces with monoatomic steps and wide terraces are obtained (Figure~\ref{fig:HF}a-d).  The $c(4\times 4)$ reconstruction is observed for such samples. If the duration of annealing at the temperature higher than 600{{\textcelsius}} is increased, SiC islanding may occur on the surface. 
Lower temperature of annealing gives rise to formation of a rough $(2\times 1)$-reconstructed surface (Figure~\ref{fig:HF}e-j).
\pb

Application of solutions based on NH$_4$F followed by any low-temperature annealing enables obtaining of clean rough Si(001) surfaces composed by narrow and short monoatomic steps (Figure~\ref{fig:SiH4F}). The $(2\times 1)$-reconstructed surface form as a result of annealing at the temperatures higher than 600{{\textcelsius}} (Figure~\ref{fig:SiH4F}d,e), $ 1\times 1$ surface was observed after treatments at lower temperatures (Figure~\ref{fig:SiH4F}g,h).
We would like to emphasize that annealings at the temperatures from 470 to 600{{\textcelsius}} result in formation of rough surfaces regardless of the applied chemical treatment (compare Figures~\ref{fig:HF}e-j and \ref{fig:SiH4F}f-h); application of solutions containing NH$_4$F always results in formation of more rough surfaces in comparison with surfaces of specimens treated in dilute HF solutions (Figure~\ref{fig:SiH4F}). 
\pb

Notice  that the $(2\times 1)$  RHEED patterns  were observed for the hydrogenated surfaces after annealing  at  800{{\textcelsius}} for 5 minutes and quenching \cite{phase_transition} which were used as the reference samples with known surface structure.
\pb
    
It should be noted also that comparison of the above STM and RHEED data makes one infer that RHEED  cannot be applied as the only method of monitoring of the surface cleaning grade and the state of dehydrated surfaces   \cite{Nanophysics-2011_H}. RHEED patterns on the hydrated surfaces corresponded to the $1\times 1$ structure. Surfaces cleaned as a result of subsequent annealings in the temperature interval from 470 to 650{{\textcelsius}} were $(1\times 1)$ or either $(2\times 1)$ or $c(4\times 4)$-reconstructed. Hence, in some cases,  a type of the RHEED pattern did not change after thermal desorption of hydrogen, however forms of the patterns, which corresponded to the $1\times 1$ structure, were different before and after annealings.
\pb

As of now, we suppose that clean Si(001) surfaces applicable for MBE formation of Ge/Si(001) heterostructures can be obtained at low temperatures (as low as 470{\textcelsius}). However, it is not excluded that further lowering of temperature of the clean Si(001) surface preparation  is possible, perhaps down to the temperatures as low as 400{{\textcelsius}} \cite{thermal_desorption}. In the latter case, Si weak flux and formation of a buffer layer may be useful to prepare a perfect enough Si(001) surface before Ge deposition.
\pb

Concluding this section, let us briefly consider the morphological peculiarities of the $c(4\times 4)$-reconstructed surface shown in Figure~\ref{fig:HF}a,b. Figure~\ref{fig:c(4x4)} presents magnified STM images of this surface.  A structure observed in the images represents a mixture of $\alpha$-$c(4\times 4)$ and $\beta$-$c(4\times 4)$ modifications \cite{Uhrberg,C-coverage}; $(2\times 1)$ and $c(4\times 4)$-reconstructed domains coexist on the surface (Figure~\ref{fig:c(4x4)}c);
location of dimers forming the $c(4\times 4)$ structure with respect to the dimers of the $(2\times 1)$ structure is also seen; ad-dimers in both epitaxial and non-epitaxial orientations are seen in (Figure~\ref{fig:c(4x4)}c).  The $\beta$-$c(4\times 4)$ modification prevails on the surface shown in Figure~\ref{fig:c(4x4)}d which is only partially occupied by $c(4\times 4)$. It is seen that the presented data are in good agreement with the model of the $c(4\times 4)$ structure proposed by Uhrberg $et\,al.$ \cite{C-coverage,Uhrberg}.
\pb

\section*{Conclusions\pb}

\subsubsection*{Is 600{{\textcelsius}} a fundamental value of temperature for Si and Ge (001) surfaces?}

Concluding the article we would like to draw the reader's attention to the fact that many of the processes described above or in the cited articles have some critical temperature close to 600{\textcelsius}. Thus, the phase transition between $2\times 1$ and $c(8\times 8)$ reconstructions occurs around this temperature \cite{phase_transition,stm-rheed-EMRS}. Exploration of dehydrogenation of the Si:H samples shows that clean surfaces obtained by annealing at the temperatures $>$\,600{{\textcelsius}} are formed by wide  terraces with monoatomic steps; the $c(4\times 4)$ reconstruction appears at these temperatures \cite{Uhrberg}. Annealings at the temperatures $<$\,600{{\textcelsius}} result in formation of rough surfaces composed by narrow and short steps. Ge QD arrays deposited by MBE at the temperatures $\apprle$ and $\apprge$\,600{{\textcelsius}} also strongly differ in both cluster compositon and nucleation. Bimodal hut arrays form at low temperatures, whereas arrays grown at high temperatures are composed by pyramids and domes. The low-temperature clusters nucleate by formation of strictly determined 2D structures composed by dimer pairs and longer chains \cite{Hut_nucleation,CMOS-compatible-EMRS}. There are two alternative scenarios of cluster formation at high temperatures: (i) similarly to the low-temperature nucleation of pyramids and (ii) by $\{$105$\}$-faceting of the Ge shapeless heaps. An assumption arises from these examples  that the temperatures do not coincide accidently, but some changes happen in the processes of migration of Si and Ge adatoms  over the (001) surface  around 600{{\textcelsius}}. 
\pb

\subsubsection*{Summary}

In summary, issues of morphology, nucleation and growth of Ge cluster arrays deposited by ultrahigh vacuum molecular beam epitaxy  on the Si(001) surface are considered in the article. 
Difference in nucleation of quantum dots during Ge deposition at  low ($\apprle$\,600{\textcelsius}) and high ($\apprge$\,600{\textcelsius}) temperatures is studied by high resolution scanning tunneling microscopy. 
 The atomic models of growth of  both species of Ge huts---pyramids and wedges---are proposed. The growth cycle of Ge QD arrays at low temperatures is explored.
 A problem of lowering of the array formation temperature is discussed with the focus on CMOS compatibility of the entire process; a special attention is paid upon approaches to reduction of treatment temperature during the Si(001) surface pre-growth cleaning, which is at once a key and the highest-temperature phase of the Ge/Si(001) quantum dot dense array formation process. The temperature of the Si clean surface preparation, the final high-temperature step of which is, as a rule, carried out directly in the MBE chamber just before the structure deposition, determines the compatibility of formation process of Ge-QD-array based devices with the CMOS manufacturing cycle. Silicon surface hydrogenation at the final stage of its wet chemical etching during the preliminary cleaning is proposed as a possible way of efficient reduction of the Si wafer pre-growth annealing temperature.

\section*{Abbreviations}
AES, Auger electron spectroscopy; 
CMOS, complementary metal-oxide semiconductor; 
CZ, Czochralski or grown by the Czochralski method; 
MBE, molecular beam epitaxy;      
ML, monolayer; PD, pairs of dimers; 
QD, quantum dot;    
RCA, Radio Corporation of America; 
RHEED, reflected high energy electron diffraction; 
RS, rebonded step; 
SIMS, secondary ion mass spectroscopy; 
STM, scanning tunneling microscope;  
XPS, X-ray photoelectron spectroscopy; 
WL, wetting layer; 
UHV, ultra-high vacuum.

\section*{Competing interests}
The authors declare that they have no competing interests.

\section*{Authors contributions}
VY conceived of the study and designed it,  processed images  and performed data analysis, took part in discussions and interpretation of the results; he also supervised  and coordinated the research projects.  
LA participated in the design of the study, carried out the experiments, performed data analysis, took part in discussions and interpretation of the results; she proposed the structural models; she also supervised the research project ($\rm \Pi$2367) and led the development of surface cleaning processes.

\section*{Acknowledgements}
  \ifthenelse{\boolean{publ}}{\small}{This research was supported by the Ministry of Education and Science of the Russian Federation under the State Contracts No.\,$ \Pi$2367, No.\,14.740.11.0052, No.\,14.740.11.0069, and  No.\,16.513.11.3046.}

\section*{Endnotes}

$^{\rm a}$The ideas underwritten  in this paragraph have already been proposed by us \cite{Sabelnik-2} and A.~V.~Dvurechenskii.

$^{\rm b}$In addition,  an  analytical chamber equipped with secondary ion mass spectroscopy (SIMS), Auger electron spectroscopy (AES) and X-ray photoelectron spectroscopy (XPS) is also available in the instrument. A Knudsen effusion cells for layer doping by boron during deposition is installed in the MBE chamber but it was not used in the described experiments.

$^{\rm c}$We express our acknowledgement to the anonymous colleague who proposed this explanation.


{\ifthenelse{\boolean{publ}}{\footnotesize}{\small}
 \bibliographystyle{bmc_article}  
  \bibliography{Ge_quantum_dot_arrays} }     


\ifthenelse{\boolean{publ}}{\end{multicols}}{}



\clearpage

\section*{Figures}

\begin{figure*}[h]
\includegraphics[scale=1]{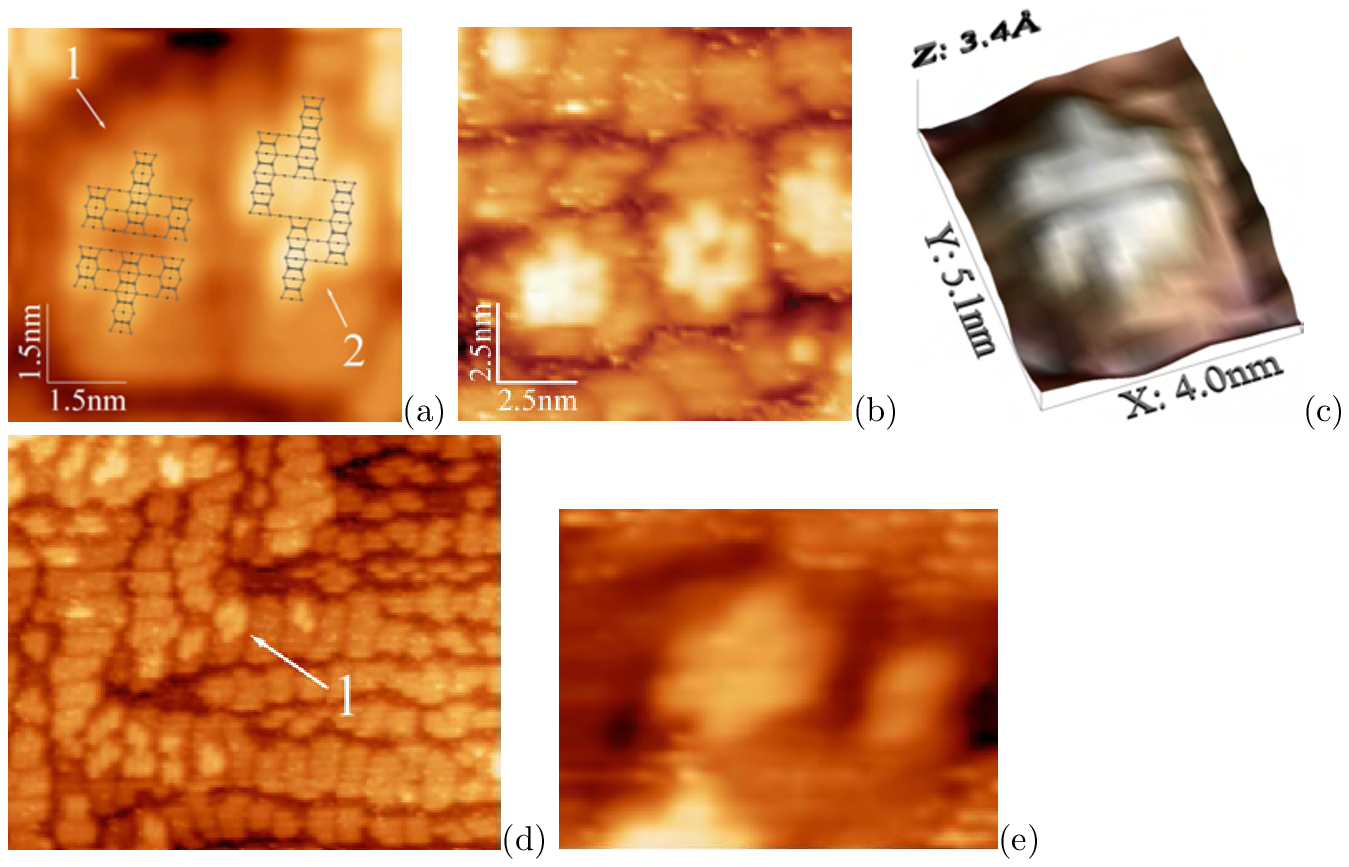}
\caption{\label{fig:nuclei}
}
\end{figure*}
\subsection*{Figure~\ref{fig:nuclei} - 
STM empty-state images of hut nuclei on Ge WL formed at different temperatures:
}
(a)--(c) pyramid (1) and wedge (2) nuclei  on the adjacent
$M \times N$ patches of WL; $T_{\rm gr} =$ 360{\textcelsius}, $h_{\rm Ge}=$ 6\,\AA; the structural models \cite{Hut_nucleation,CMOS-compatible-EMRS} are superimposed on the corresponding images in (a);
(b),(c) pyramid nuclei on WL formed at low temperature ($T_{\rm gr} =$ 360{\textcelsius}): 
(b)  $h_{\rm Ge}=$ 5.4\,\AA;
(c) $h_{\rm Ge}=$ 6\,\AA;
(d),(e) a pyramid nucleus on WL formed at high temperature, $h_{\rm Ge}=$ 5\,\AA:
(d) $T_{\rm gr} =$ 600{\textcelsius}, $43\times 37$\,nm;
(e) $T_{\rm gr} =$ 650{\textcelsius},  $7.8\times 6$\,nm.

\clearpage
\begin{figure*}[h]
\includegraphics[scale=1]{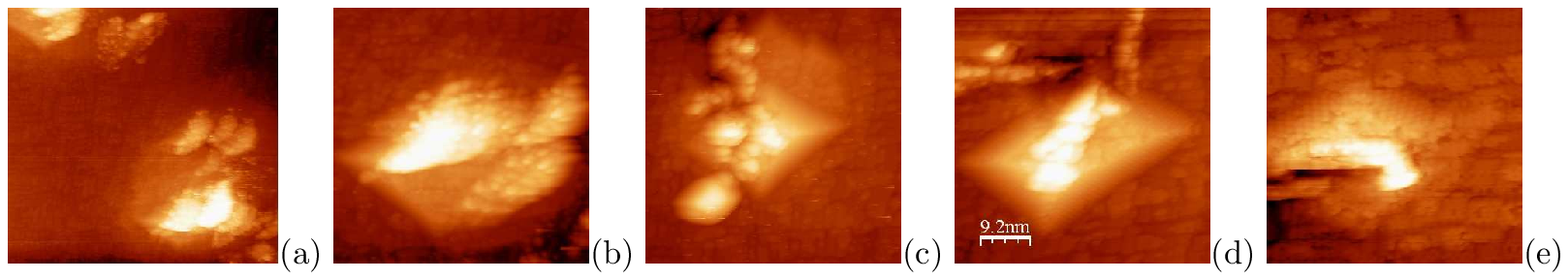}
\caption{\label{fig:heap}
}
\end{figure*}
\subsection*{Figure~\ref{fig:heap} - 
Formation of $\{$105$\}$ facets from shapeless areas with excess of Ge:
}
STM images of different phases of faceting, $T_{\rm gr} =$ 650{\textcelsius}, $h_{\rm Ge}=$ 5\,\AA;
(a) a shapeless Ge `heap' without faceting, $150\times 141$\,nm;
(b) at the outset of faceting, $64\times 64$\,nm;
(c)--(d) after growth stoppage and annealing at the growth temperature;
(c) $72\times 72$\,nm;
(d) $46\times 46$\,nm;
(e) $23\times 23$\,nm.
\pb

\clearpage
\begin{figure*}
\includegraphics[scale=1]{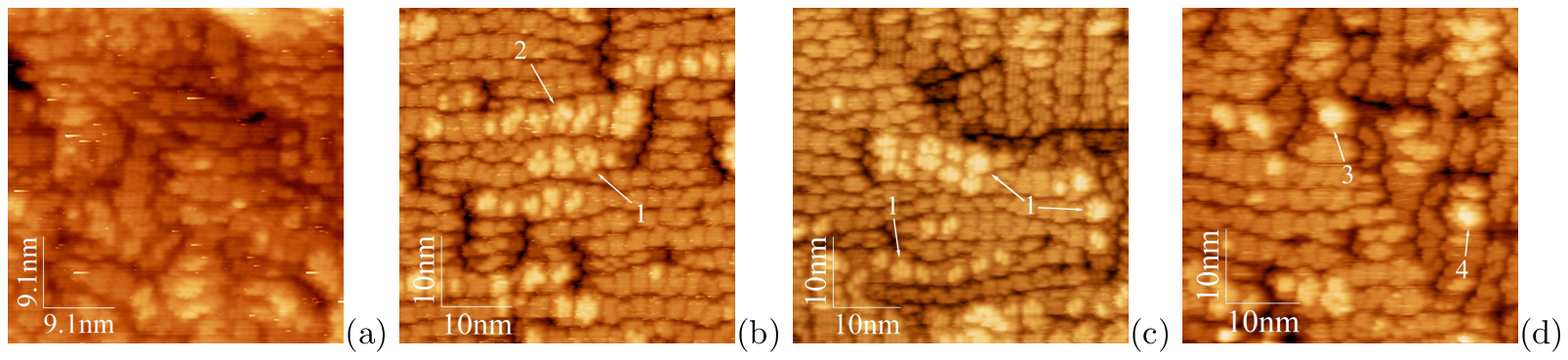}
\caption{\label{fig:4,4A_5,1A}
}
\end{figure*}
\subsection*{Figure~\ref{fig:4,4A_5,1A} - 
STM images of  Ge  WL on Si(001) at the outset of QD array formation: 
}
$T_{\rm gr} = 360${\textcelsius},
(a) $h_{\rm Ge}=4.4$~\r{A},  $U_{\rm s}=-1.86$~V, $I_{\rm t}=100$~pA, neither hut clusters nor their nuclei are observed;  
(b) $h_{\rm Ge}=5.1$~\r{A},  $U_{\rm s}=+1.73$~V, $I_{\rm t}=150$~pA; 
(c) $U_{\rm s}=+1.80$~V, $I_{\rm t}=100$~pA; 
(d) $U_{\rm s}=+2.00$~V, $I_{\rm t}=100$~pA. 
Examples of characteristic features are numbered as follows:  nuclei of pyramids (1) and  wedges  (2) [1\,ML high over WL patchs, Figure~\ref{fig:nuclei}],\cite{Hut_nucleation,CMOS-compatible-EMRS}  small pyramids (3) and wedges (4) [2\,ML high over WL patchs]\cite{Hut_nucleation,CMOS-compatible-EMRS,classification,atomic_structure}.
\pb

\clearpage
\begin{figure*}
\includegraphics[scale=1]{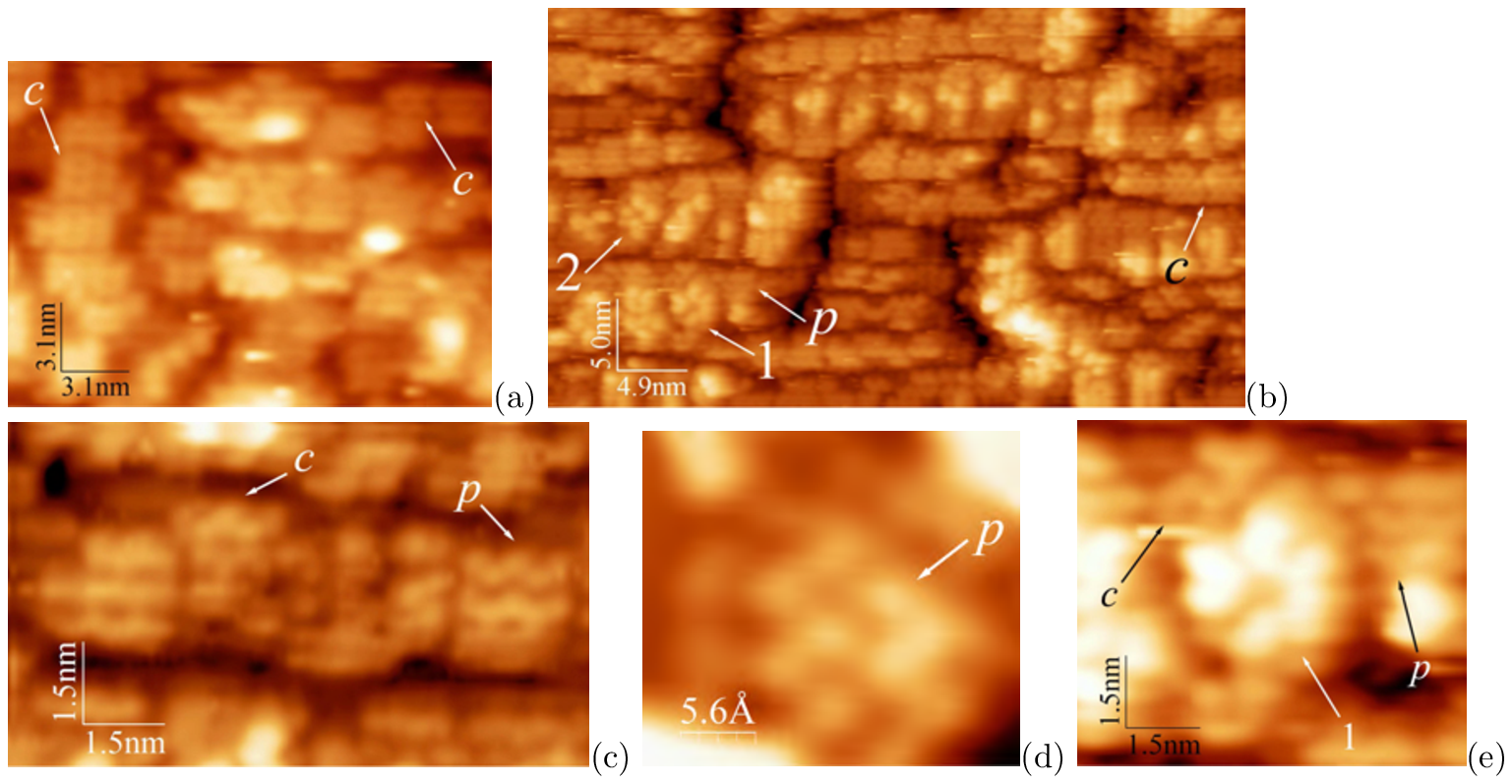}
\caption{\label{fig:WL_patches}
}
\end{figure*}
\subsection*{Figure~\ref{fig:WL_patches} - 
STM images of  Ge WL on Si(001): 
}
$T_{\rm gr} =$ 360{\textcelsius},
the ordinary $c(4\times 2)$~$(c)$  and $p(2\times 2)$~$(p)$ reconstructions within the $M\times N$ patches are often observed simultaneously,  
(a) $h_{\rm Ge}=4.4$~\r{A}, $U_{\rm s}=-1.86$~V, $I_{\rm t}=100$~pA, 
 only the $c(4\times 2)$ structure is resolved; 
(b)~$h_{\rm Ge}=5.1$~\r{A}, $U_{\rm s}=-3.78$~V, $I_{\rm t}=100$~pA, 
 both $c(4\times 2)$  and $p(2\times 2)$ structures are revealed as well as  nuclei of a pyramid (1) and a wedge (2); 
(c),(d) $h_{\rm Ge}=6.0$~\r{A}, $U_{\rm s}=+1.80$~V, $I_{\rm t}=80$~pA, both $c(4\times 2)$  and $p(2\times 2)$ reconstructions are well resolved; 
(e) $h_{\rm Ge}=5.1$~\r{A}, $U_{\rm s}=-3.78$~V, $I_{\rm t}=100$~pA, a pyramid nucleus  on the $c(4\times 2)$ reconstructed patch with the adjacent  $p(2\times 2)$  reconstructed patch.
\pb

\clearpage
\begin{figure*}[h]
\includegraphics[scale=1]{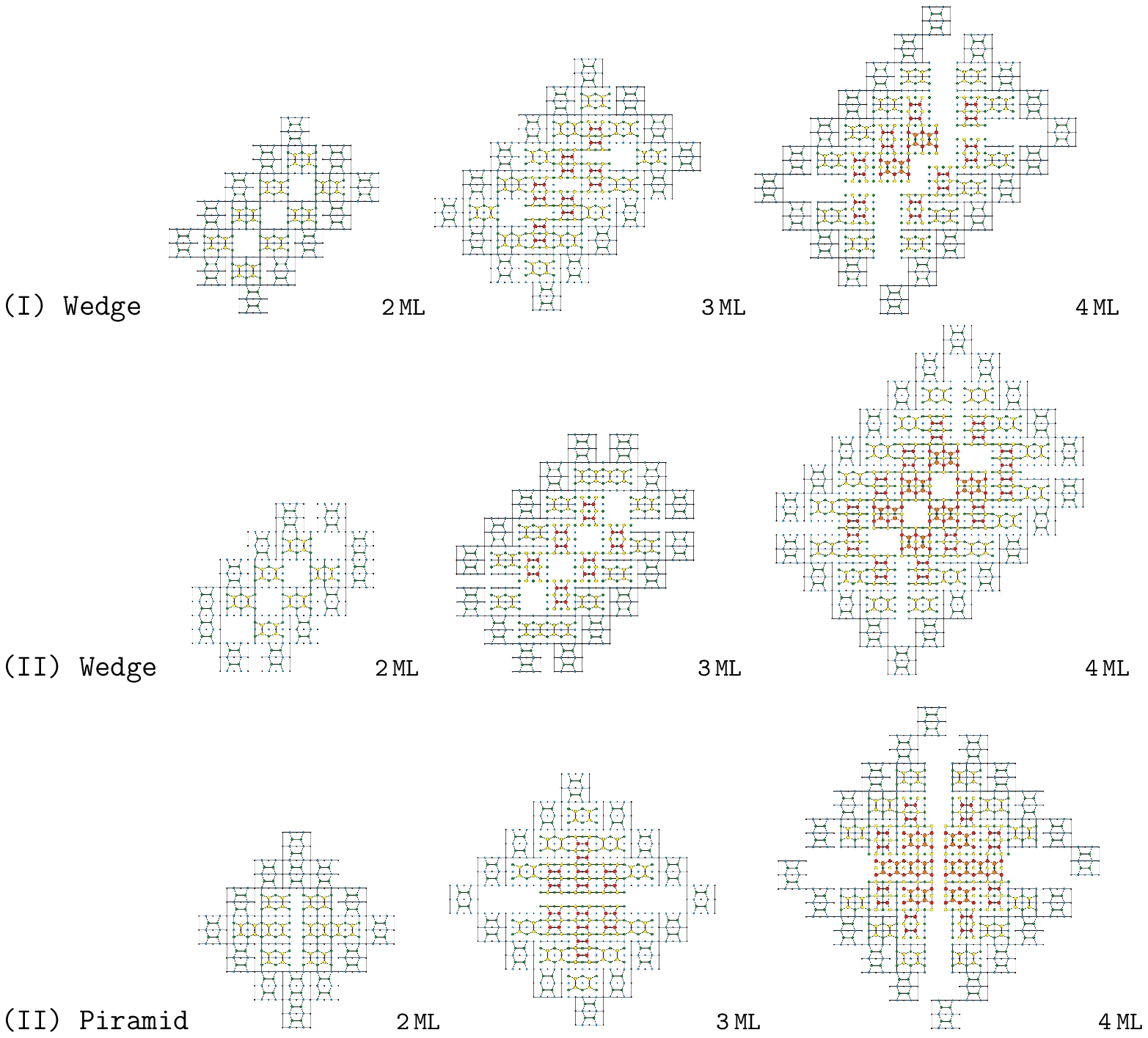}
\caption{\label{fig:growth}
}
\end{figure*}
\subsection*{Figure~\ref{fig:growth} - 
Models of Ge hut growth:
}
 (I) uniform  addition of Ge atoms to four facets;
 (II) nun-uniform addition of Ge atoms to facets.
\pb

\clearpage
\begin{figure*}[h]
\includegraphics[scale=1]{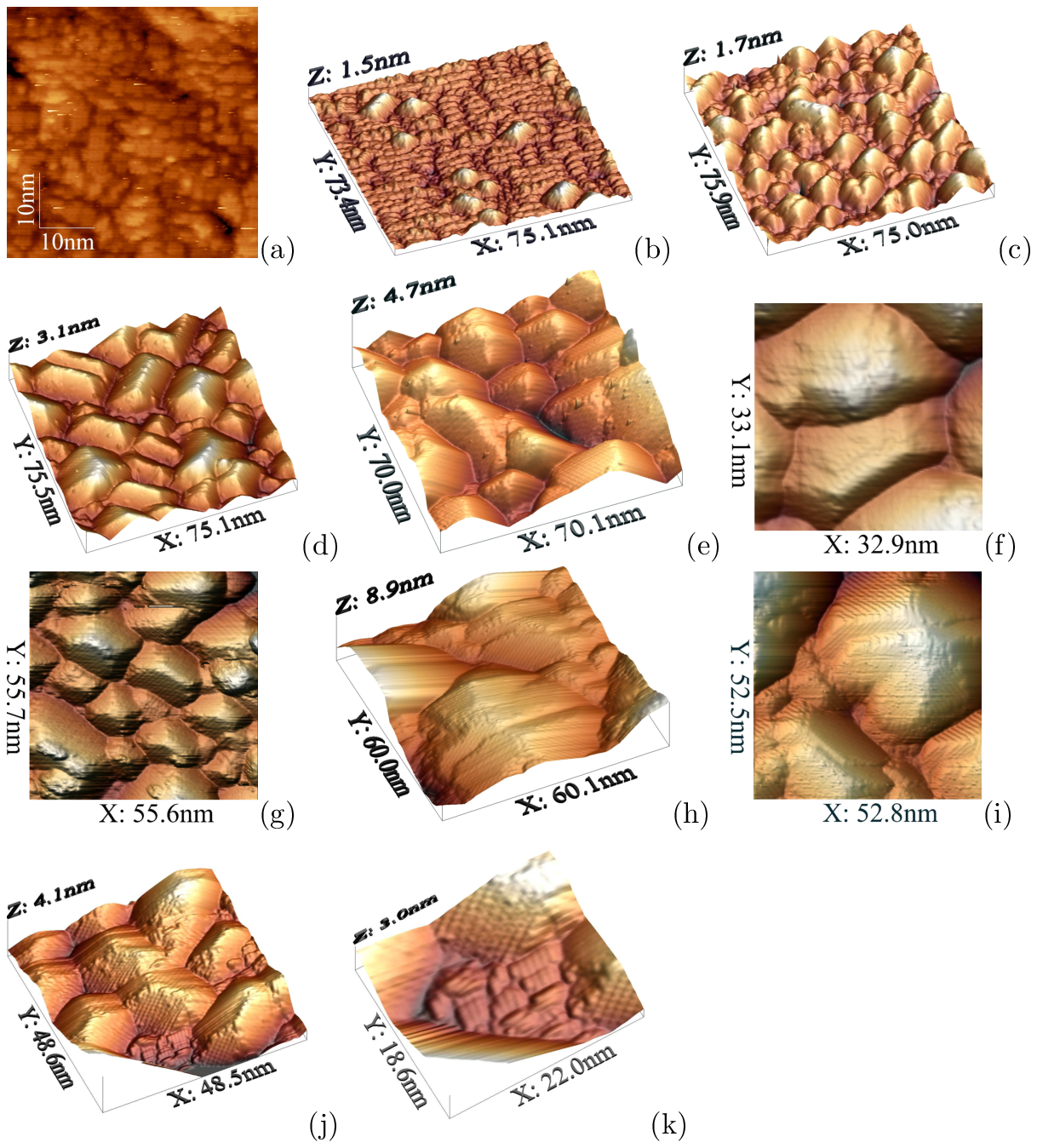}
\caption{\label{fig:array_life}
}
\end{figure*}
\subsection*{Figure~\ref{fig:array_life} - 
STM images of a Ge QD dense array at different phases of its evolution from patched WL to 2D nanocrystalline layer: 
}
$T_{\rm gr} =$ 360{\textcelsius}, $h_{\rm Ge}=$ 
(a) 4.4 \AA, before nucleation (see also Figure~\ref{fig:4,4A_5,1A} for details of array nucleation at $h_{\rm Ge}=5.1$ \AA); 
(b) 6 \AA, growing small huts, nucleation goes on; 
(c) 8 \AA, maximum density ($\sim\,6\times 10^{11}$\,cm$^{-2}$); 
(d) 10 \AA, maximum uniformity, large huts start to coalesce; 
(e),(f) 14 \AA, huts go on coalescing; 
(g) 15 \AA, 2D layer starts to form;
(h),(i),(j),(k) 18 \AA, 2D nanocrystallyne film grows, chaos of faceted hillocks and pits (i) is observed;
however, Ge WL $(M\times N)$-patched structure is clearly resolved on bottom of pits (j),(k).
\pb

\clearpage
\begin{figure*}[h]
\includegraphics[scale=1]{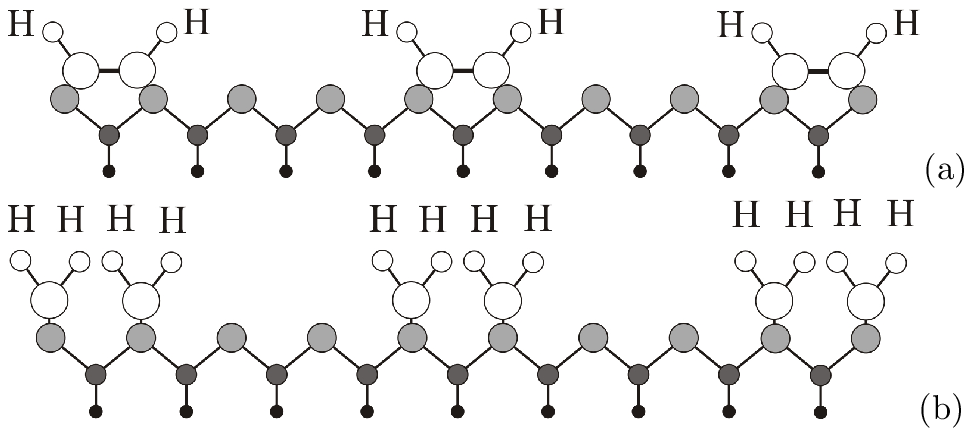}
\caption{\label{fig:SiH_schematic}
}
\end{figure*}
\subsection*{Figure~\ref{fig:SiH_schematic} - 
Structure of a hydrated Si(001) surface:
}
(a) mono-hydride and (b) di-hydride.

\clearpage
\begin{figure*}[h]
\includegraphics[scale=1]{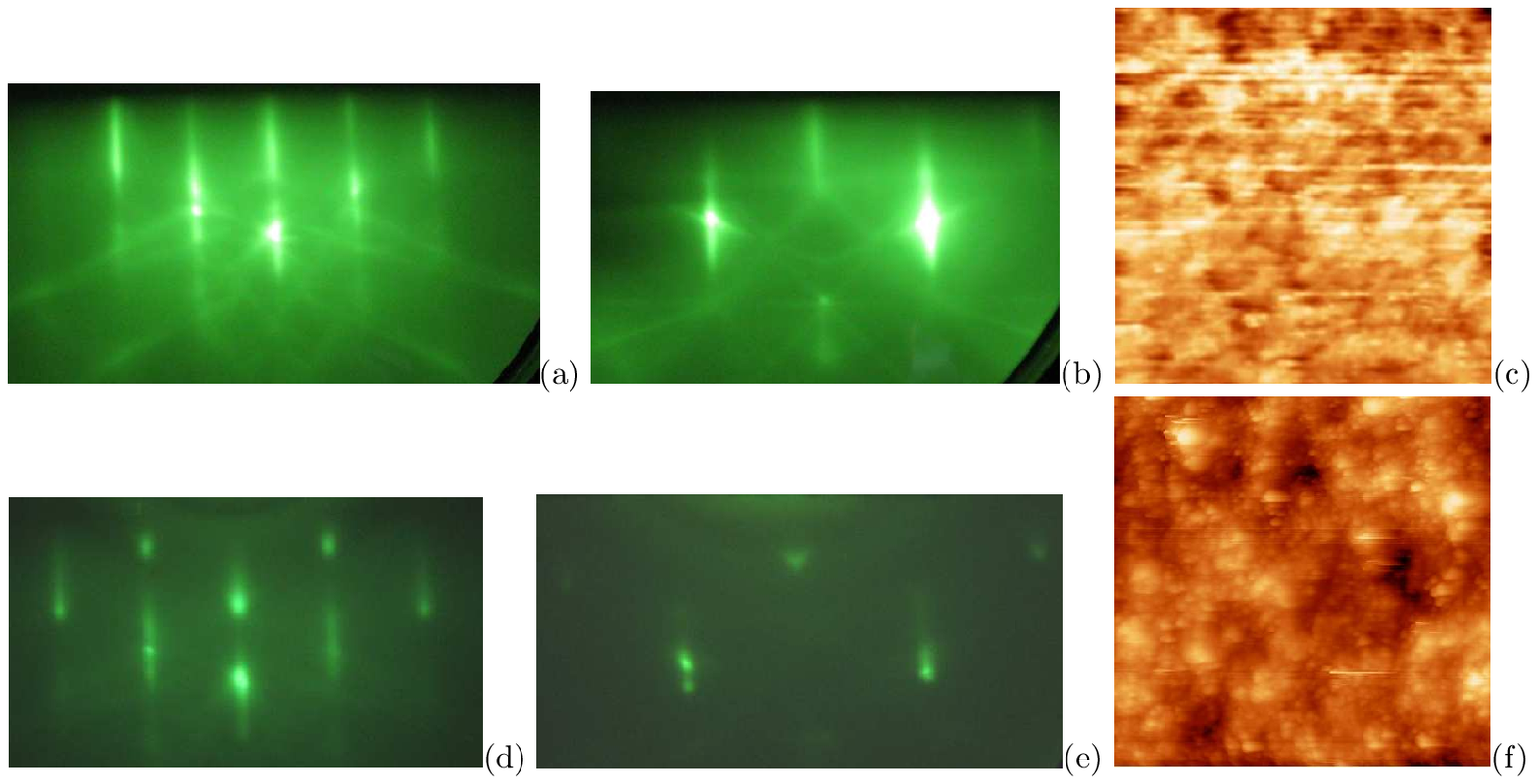}
\caption{\label{fig:SiH}
}\end{figure*}
\subsection*{Figure~\ref{fig:SiH} - 
RHEED patterns and STM images of Si:H surfaces obtained as a result of different chemical treatments:
}
(a)--(c) after hydrogenation in dilute HF; 
(d)--(f) after hydrogenation in buffered HF\,+\,NH$_4$F; 
RHEED patterns: $E = 10$\,keV, 
(a),(d) [110] azimuth, 
(b),(e) [010] azimuth; 
STM empty-state  images: 
(c) $100\times 100$\,nm, $U_{\rm s}=+1.9$~V, $I_{\rm t}=100$~pA; 
(f) $88\times 88$\,nm, $U_{\rm s}=+2.0$~V, $I_{\rm t}=100$~pA.

\clearpage
{\begin{figure*}[h]
\includegraphics[scale=1]{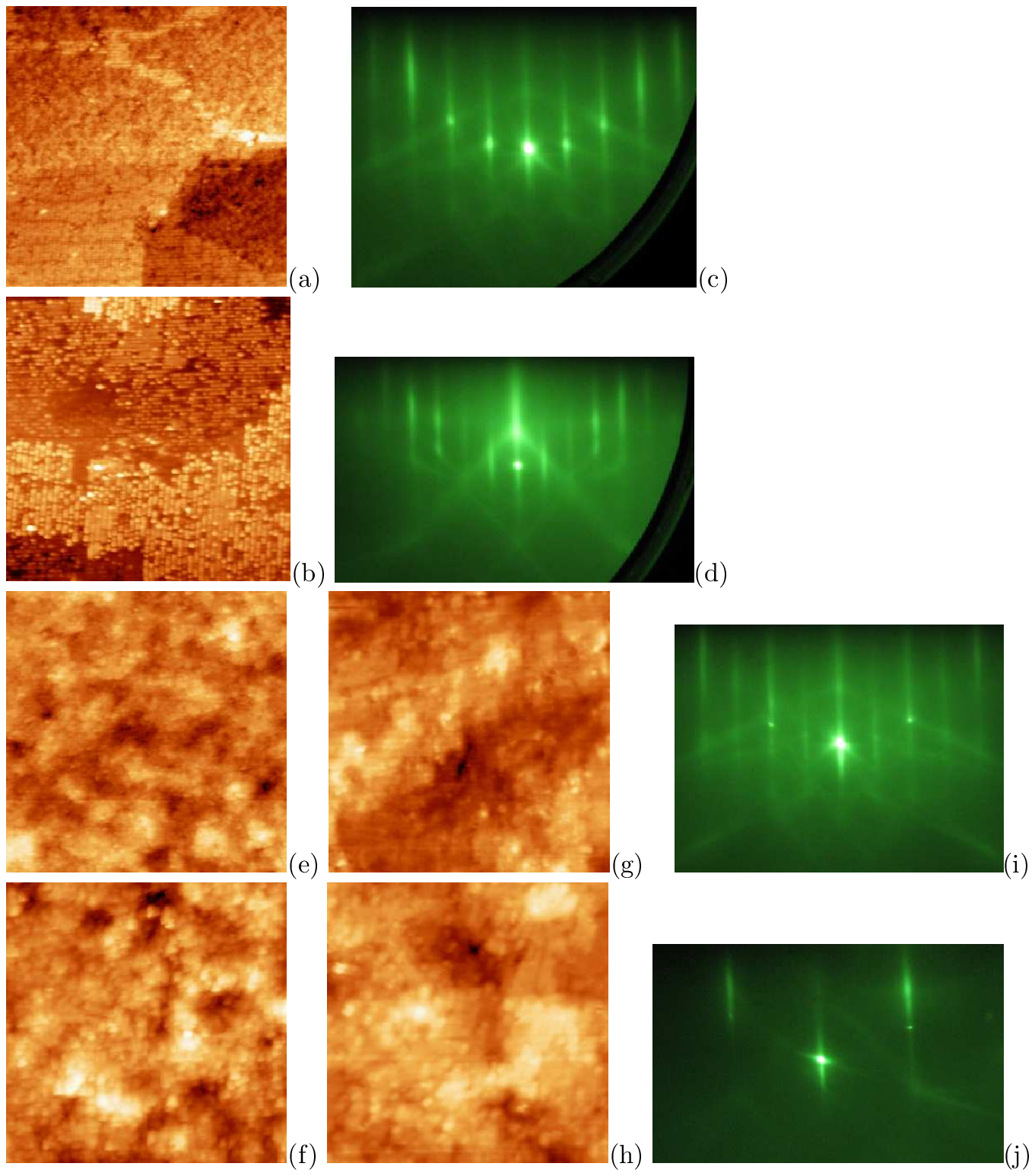}
\caption{\label{fig:HF}
}
\end{figure*}
\subsection*{Figure~\ref{fig:HF} - 
STM images and RHEED patterns of Si:H surfaces obtained as a result of hydrogenation in dilute HF after different heat treatments:
}
(a),(b),(e)--(h) STM empty-state images;
(a) 650{{\textcelsius}} for 8\,min, $57\times 57$\,nm;
(b) 610{{\textcelsius}} for 10\,min, $41\times 41$\,nm;
(c),(d) corresponding RHEED patterns, $E = 10$\,keV:
(c) [110],
(d) [010];
(e) 570{{\textcelsius}} for 20\,min, $101\times 101$\,nm;
(f) 550{{\textcelsius}} for 30\,min, $66\times 66$\,nm;
(g) 530{{\textcelsius}} for 35\,min, $41\times 41$\,nm;
(h) 500{{\textcelsius}} for 35\,min, $49\times 49$\,nm;
(i),(j) corresponding RHEED patterns, $E = 10$\,keV:
(i) [110],
(j) [010].
}

\clearpage
{\begin{figure*}[h]
\includegraphics[scale=1]{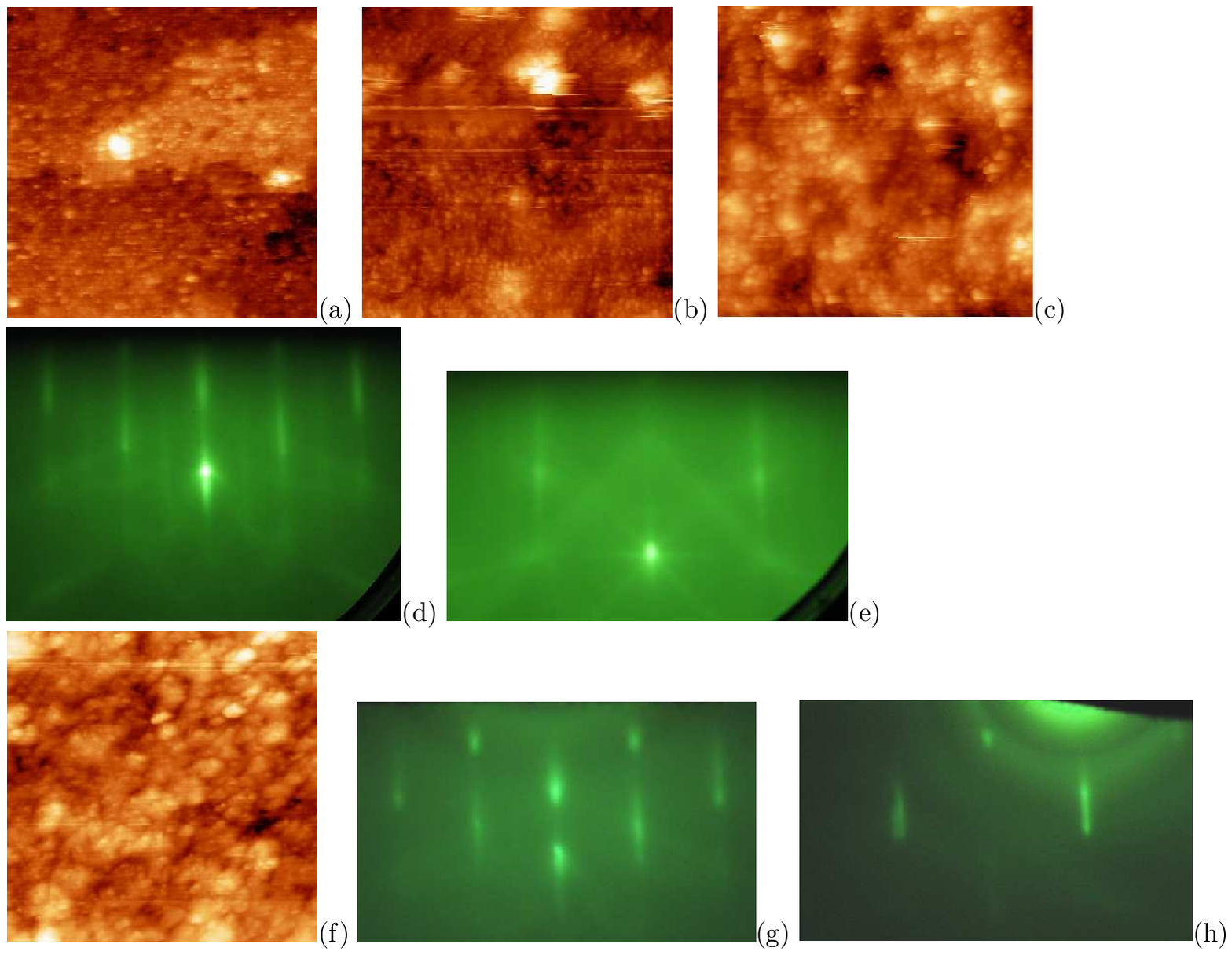}
\caption{\label{fig:SiH4F}
}
\end{figure*}
\subsection*{Figure~\ref{fig:SiH4F} - 
STM images and RHEED patterns of Si:H surfaces obtained as a result of hydrogenation in NH$_4$F or HF\,+\,NH$_4$F solution after different heat treatments:
}
(a)--(c),(f) STM empty-state images;
(a) NH$_4$F, 650{{\textcelsius}} for 5\,min, $40\times 40$\,nm;
(b) HF\,+\,NH$_4$F, 610{{\textcelsius}} for 10\,min, $56\times 56$\,nm;
(c) NH$_4$F, 610{{\textcelsius}} for 10\,min, $88\times 87$\,nm;
(d),(e) corresponding RHEED patterns, $E = 10$\,keV:
(d) [110],
(e) [010];
(f) NH$_4$F, 550{{\textcelsius}} for 35\,min, $60\times 60$\,nm;
(g),(h) corresponding RHEED patterns, $E = 10$\,keV:
(g) [110],
(h) [010].
}

\clearpage
{\begin{figure*}[h]
\includegraphics[scale=.6]{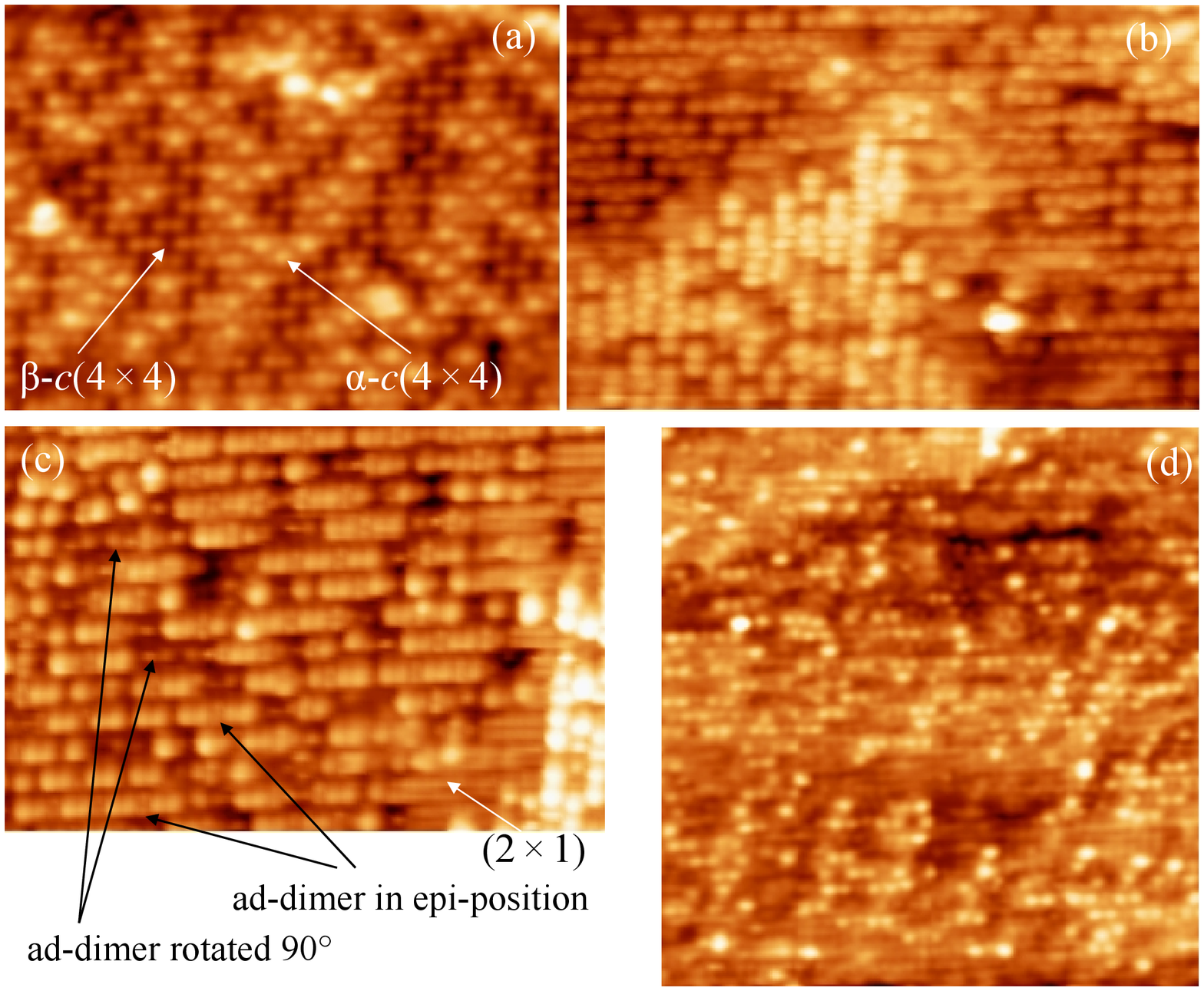}
\caption{\label{fig:c(4x4)}
}
\end{figure*}
\subsection*{Figure~\ref{fig:c(4x4)} - 
STM images of the Si(001)-$c(4\times 4)$ surface:
}
(a) empty states,  $19\times 14$\,nm, $U_{\rm s}=+2.5$~V, $I_{\rm t}=120$~pA;
(b) empty states,  $19\times 12$\,nm, $U_{\rm s}=+2.0$~V, $I_{\rm t}=120$~pA;
(c) filled states,  $22\times 15$\,nm, $U_{\rm s}=-3.9$~V, $I_{\rm t}=150$~pA;
(d)  filled states, $19\times 20$\,nm,   $U_{\rm s}=-3.9$~V, $I_{\rm t}=150$~pA.  As it is clearly observed in (a), the structure is composed by a mixture of the $\alpha$-c$(4\times 4)$ and $\beta$-c$(4\times 4)$ modifications; it is seen in (c) that the $c(4\times 4)$ and $(2\times 1)$ reconstructions coexist on the surface;  location of dimers forming the $c(4\times 4)$ structure with respect to the dimers of the $(2\times 1)$ structure is also seen; ad-dimers in both epitaxial and non-epitaxial orientations are seen in (c).  The $\beta$-$c(4\times 4)$ modification prevails in (d) which is only partially occupied by $c(4\times 4)$.
}
\pb

\end{bmcformat}
\end{document}